\documentclass[a4paper,12pt]{article}
\usepackage[utf8]{inputenc}

\usepackage[left=2.5cm,right=2.5cm,bottom=4cm]{geometry}

\usepackage{mathtools,amsmath,amssymb}
\usepackage{graphicx}
\usepackage{comment}
\usepackage[english]{babel}

\numberwithin{equation}{section} 

\usepackage{lmodern}
\usepackage[T1]{fontenc}
\usepackage{microtype}

\usepackage{hyperref}

\usepackage{tikz}

\newcommand{\NN}{\mathbf{N}}
\newcommand{\half}{{\frac{1}{2}}}

\newcommand{\mon}{\mathcal{M}}
\newcommand{\bmon}{\hat{\mathcal{M}}}
\newcommand{\qmon}{\mathrm{M}}
\newcommand{\qh}{\mathrm{U}}
\newcommand{\trr}{\tr_0}
\newcommand{\cc}{\mathcal{Y}}
\newcommand{\OO}{O}
\newcommand{\osc}[1]{\boldsymbol{\mathrm{#1}}}

\newcommand{\oa}{\osc{a}}

\newcommand{\oad}{\osc{\bar{a}}}

\newcommand{\LL}{L}
\newcommand{\plax}{\mathbf{L}}

\newcommand{\qop}{\mathbf{Q}}

\renewcommand{\top}{\mathbf{T}}
\newcommand{\KQL}{\mathrm{K}}
\newcommand{\KQR}{\hat{\mathrm{K}}}
\newcommand{\KL}{\mathcal{K}}
\newcommand{\KR}{\hat{\mathcal{K}}}
\newcommand{\rtran}{\mathcal{U}}

\newcommand{\rfund}{R}
\newcommand{\perm}{\mathbb{P}}
\newcommand{\set}{I}

\renewcommand{\a}{0}
\newcommand{\loz}{z\gg 1}
\newcommand{\br}{q}
\newcommand{\bl}{p}

\DeclareMathOperator{\tr}{tr}
\DeclareMathOperator{\Li}{Li}

\begin{document}

\begin{titlepage}\strut\hfill DCPT-15/55

\vspace{1in}

\begin{center}
 \textbf{\Large Q-operators for the open Heisenberg spin chain}
\\\vspace{1in}
{\large Rouven Frassek and István M. Szécsényi}
\\[0.2in]

 Department of Mathematical Sciences, Durham University,\\
     South Road, Durham DH1 3LE, United Kingdom
\\[0.6in]
 \texttt{$\big\{$rouven.frassek,i.m.szecsenyi$\big\}$@durham.ac.uk
    }
    \\[1cm]
    
 \end{center}
 \vspace{.8in}
\begin{center}
\textbf{\large Abstract}
\end{center}
\begin{center}
\begin{minipage}{400pt}
 \noindent  
 We construct Q-operators for the open spin-$\frac{1}{2}$ {\small XXX} Heisenberg spin chain with diagonal boundary matrices. 
The Q-operators are defined as traces over an infinite-dimensional auxiliary space involving novel types of reflection operators derived from the boundary Yang-Baxter equation. We argue that the Q-operators defined in this way are polynomials in the spectral parameter and show that they commute with transfer matrix. Finally, we prove that the Q-operators satisfy Baxter's TQ-equation and derive the explicit form of their eigenvalues in terms of the Bethe roots. 
\end{minipage}
\end{center}
\end{titlepage}

\vfill
\newpage
\tableofcontents
\vspace{2cm}
\section{Introduction}\label{sec:intro}
Integrable spin chains are prominent examples of quantum integrable models. They bear a close relation to two-dimensional integrable quantum field theories as well as lattice models and rest on surprisingly rich mathematical structures. Furthermore, integrable spin chains play an important role in condensed matter physics and in the {\small AdS/CFT} correspondence.

The key example of an integrable spin chain is the periodic spin-$\frac{1}{2}$ {\small XXX} Heisenberg spin chain. This one-dimensional quantum system describes the interaction of neighbouring spins aligned on a line with closed periodic boundary conditions and has been solved by Hans Bethe \cite{Bethe1931}.

Curiously, the spin chain need not have closed boundary conditions to be integrable. In fact, it was shown in \cite{Alcaraz1987} that the nearest-neighbour Hamiltonian for the Heisenberg spin chain with open boundary conditions
\begin{equation}\label{eq:hamiltonian}
 \mathbf{H}=\sum_{i=1}^{\LL-1} \vec \sigma_i\cdot \vec \sigma_{i+1} +\xi\, \sigma_1^3+\hat\xi\,\sigma_\LL^3\,,
\end{equation} 
can be solved using the coordinate Bethe ansatz. 
Here $\xi$ and $\hat \xi$ are arbitrary boundary parameters, $\LL$ denotes the length of the spin chain and $\vec\sigma_i$ the Pauli matrices at spin chain site $i$ acting on the $2^\LL$-dimensional quantum space.

The quantum inverse scattering method is a powerful tool to study integrable models. It allows the transfer matrix to be constructed using R-matrices that are solutions of the Yang-Baxter equation. An introduction, as well as references, can be found in \cite{Faddeev2007}. Initially formulated for closed systems, the quantum inverse scattering method was further extended to integrable models with open boundaries by Sklyanin \cite{Sklyanin1988}. In particular, this allowed the transfer matrix to be constructed for the open {\small XXX} Heisenberg spin chain. As for the  closed case the method relies on the Yang-Baxter equation but in addition to the R-matrices also incorporates certain boundary matrices obeying the boundary Yang-Baxter equation. \footnote{Throughout this article we restrict ourselves to the case of diagonal boundary matrices with the Hamiltonian given in \eqref{eq:hamiltonian}.}

Baxter Q-operators are of distinguished importance in the theory of integrable systems. They contain the information about the eigenfunctions as well as the Bethe roots. They are named after Baxter who introduced a Q-operator to solve the eight-vertex model \cite{Baxter2007}. Nowadays, interest in the Q-operators stems from various branches of theoretical physics. Q-operators are of central importance in the {\small ODE/IM} correspondence \cite{DoreyJ.Phys.A40:R2052007} which relates quantum integrable models to ordinary differential equations and play a major role in the formulation of the fermionic basis \cite{Boos2008}. A systematic, representation theoretical approach to Q-operators was recently studied in \cite{Hernandez2011,Frenkel2013}. Furthermore, the results obtained for the spectral problem of planar $\mathcal{N}=4$ super Yang-Mills theory, see \cite{Gromov2014a} and references therein, suggest that Q-operators play a central role in the study of the integrable structures appearing in the context of the {\small AdS/CFT} correspondence, see e.g.~\cite{Beisert2010a} and references therein and thereof. 

Even though the Q-operator is often referred to as the most powerful tool for the exact diagonalisation of integrable systems it has only been studied for a few examples. In fact, the construction for the periodic {\small XXX} Heisenberg chain in the presence of a magnetic field, along the lines of \cite{Bazhanov1999}, is rather recent \cite{Bazhanov2010a}. See also \cite{Frassek2015}, where these Q-operators have been diagonalised using the algebraic Bethe ansatz. Apart from the construction in terms of integral operators for certain non-compact open magnets that in particular appeared in the context of {\small QCD} \cite{DerkachovJHEP0310:0532003,Derkachov2006} little is known about Q-operators for open spin chains.

In this article we construct the Q-operators for the open {\small XXX} spin-$\frac{1}{2}$ Heisenberg spin chain. Following the quantum inverse scattering method we construct the Q-operators as traces of single-row monodromies similar to the ones that appeared in \cite{Bazhanov2010a} and novel types of boundary operators which are derived from the boundary Yang-Baxter equation. We show that the Q-operators commute with the transfer matrix, prove Baxter's TQ-equation on the operatorial level and derive their eigenfunctions.

The article is organised as follows. We first review the construction of the transfer matrix for the open {\small XXX} spin-$\frac{1}{2}$ Heisenberg spin chain in Section~\ref{sec:openchain}. In Section~\ref{sec:qop} we define the Q-operators. We introduce the Lax operators for the single-row monodromies and derive the boundary operators for the Q-operators from the boundary Yang-Baxter equation. Section~\ref{sec:qprops} discusses various properties of the two Q-operators. We show that the transfer matrix commutes with the Q-operators and elaborate on the symmetry which relates them. Furthermore, we carefully study the trace in the auxiliary space of the Q-operators and show that it yields polynomials in the spectral parameter. More precisely, we demonstrate that the entries of the Q-operator are arranged in blocks. We evaluate them at leading order of the spectral parameter. In Section~\ref{sec:tq} we prove Baxter's TQ-equation on the operatorial level and argue that the eigenfunctions of the Q-operators yield the desired Q-functions. Finally, we discuss the results and provide an outlook in Section~\ref{sec:conc}.

\section{The open Heisenberg spin chain}\label{sec:openchain}
In this section we review the construction of the fundamental transfer matrix of the open {\small XXX} Heisenberg spin chain in the framework of the quantum inverse scattering method following \cite{Sklyanin1988}.

At first we define the fundamental R-matrix
\begin{equation}\label{eq:rmatrix}
 \rfund(z)=z+\perm\,,
\end{equation} 
with the permutation matrix $\perm=\sum_{a,b=1}^2 e_{ab}\otimes e_{ba}$. Here $z$ is the spectral parameter and $e_{ab}$ denotes the $2\times 2$ matrices with the entry ``$1$'' in the $a^\text{th}$ row and $b^\text{th}$  column and ``$0$'' elsewhere. Thus, the R-matrix acts on the tensor product of two spaces 
\begin{equation}\label{eq:ract}
 R(z):\mathbb{C}^2\otimes\mathbb{C}^2\rightarrow\mathbb{C}^2\otimes\mathbb{C}^2\,.
\end{equation} 
The R-matrix in \eqref{eq:rmatrix} is a solution to the Yang-Baxter equation
\begin{equation}\label{eq:rrr}
 R_{12}(x-y)R_{13}(x-z)R_{23}(y-z)=R_{23}(y-z)R_{13}(x-z)R_{12}(x-y)\,,
\end{equation} 
where we defined
\begin{equation}
 R_{12}(z)=R(z)\otimes \mathbb{I}\,,
\end{equation} 
and $R_{13}$, $R_{23}$, correspondingly. Furthermore, one finds that the R-matrix satisfies the unitarity and crossing unitarity relation
\begin{equation}\label{eq:unitR}
 R(z)R(-z)=1-z^2\,,\qquad R^{t_1}(z)R^{t_1}(-z-2)=-z(z+2)\,.
\end{equation} 
Here $t_1$ denotes the transposition in the first space of the symmetric R-matrix, cf.~\eqref{eq:ract}.

In order to construct the transfer matrix of the open spin chain we define the quantum Lax operators $\mathcal{L}^{[i]}$ which coincide with the R-matrices introduced above 
\begin{equation}\label{eq:tlax}
\mathcal{L}^{[i]}(z)\equiv R_{0,i}(z)\,.
\end{equation} 
They act on the $i^\text{th}$ spin chain site in the quantum space and on the two-dimensional auxiliary space denoted by ``$0$'' which has been suppressed in the notation above. The transfer matrix $\top$ of the open spin chain is defined from the double-row monodromy matrix $\mathcal{U}$ as 
\begin{equation}\label{eq:trans}
 \top(z)=\trr \KL (z)\mathcal{U}(z)\,\qquad\text{with}\qquad  \mathcal{U}(z)=\mon(z)\KR (z)\bmon(z)\,.
\end{equation}
The single-row monodromy matrices $\mon(z)$ and $\bmon(z)$ are constructed from the tensor product of the Lax operators $\mathcal{L}^{[i]}$ with $i=1,\ldots,\LL$ in the quantum space and multiplication in the auxiliary space as
\begin{equation}\label{eq:mon}
 \mon(z)=\mathcal{L}^{[1]}(z)\mathcal{L}^{[2]}(z)\cdots \mathcal{L}^{[\LL]}(z)\,,\quad\quad  \bmon(z)=\mathcal{L}^{[\LL]}(z)\mathcal{L}^{[\LL-1]}(z)\cdots \mathcal{L}^{[1]}(z)\,.
\end{equation} 
Here $\LL$ is the length of the spin chain, i.e. the number of sites that build up the $2^\LL$-dimensional quantum space.
The boundary operators $ \KL $ and  $\KR$ are diagonal matrices that only act non-trivially in the auxiliary space and read
\begin{equation}\label{eq:kmatrices}
 \KL (z)=\left(\begin{array}{cc}
              \bl +z+1&0\\
	      0&\bl-z-1
             \end{array}\right)\,,\quad\quad  \KR (z)=\left(\begin{array}{cc}
              \br +z&0\\
	      0&\br-z
             \end{array}\right)\,.
\end{equation} 
Here, as discussed at the end of this section, the boundary parameters in \eqref{eq:hamiltonian} are related to $p$ and $q$ via $\xi=1/p$ and $\hat\xi=1/q$. 
The boundary operators in \eqref{eq:kmatrices} can be determined from the boundary Yang-Baxter equation. In particular, the following boundary Yang-Baxter equations hold
\begin{equation}\label{bybef3}
 \rfund(x-y)(\rtran(x)\otimes \mathbb{I}) \rfund(x+y)(\mathbb{I}\otimes \rtran(y) )=(\mathbb{I}\otimes \rtran(y)) \rfund(x+y)(\rtran(x)\otimes \mathbb{I}) \rfund(x-y)\,,
\end{equation} 
and
\begin{equation}\label{bybef2}
(\mathbb{I}\otimes \KL (y)) \rfund(-x-y-2)(\KL (x)\otimes \mathbb{I}) \rfund(y-x)= \rfund(y-x)(\KL (x)\otimes \mathbb{I}) \rfund(-x-y-2)(\mathbb{I}\otimes \KL (y) )\,,
\end{equation} 
where in addition to the explicit form of the boundary matrices \eqref{eq:kmatrices} we used the Yang-Baxter equation \eqref{eq:rrr} to show \eqref{bybef3} which involves the double-row monodromy $\mathcal{U}$. 

The transfer matrix satisfies the crossing relation
\begin{equation}\label{eq:tcros}
 \top(z)=\top(-z-1)\,.
\end{equation} 
This relation is shown in Appendix~\ref{app:tcros} as it is needed in Section~\ref{sec:tq}. For alternative proofs see e.g.~\cite{MezincescuNucl.Phys.B372:597-6211992,Behrend1996}.
As a consequence of the boundary Yang-Baxter equations above and the unitarity relations in \eqref{eq:unitR} one finds that the transfer matrix commutes for different values of the spectral parameter
\begin{equation}
 [ \top(x), \top(y)]=0\,.
\end{equation} 
The spectrum of the transfer matrix can be determined using Bethe ansatz methods. The eigenvalues of the transfer matrix $T$ are given in terms of Baxter Q-functions by the TQ-equation
\begin{equation}\label{eq:baxtereqn}
\begin{split}
 T(z)=(z\pm p)(z\pm q)\frac{(z+1)^{2\LL+1}}{z+\half}\frac{Q_\pm(z-1)}{Q_\pm(z)}+(z\mp p+1) (z\mp q+1)\frac{z^{2\LL+1}}{z+\half}\frac{Q_\pm(z+1)}{Q_\pm(z)}\,.
 \end{split}
\end{equation}
The TQ-equation can be obtained from the algebraic Bethe ansatz. Depending on the choice of the reference state, all spin up or all spin down, the eigenvalues of the transfer matrix are either parametrised by the Q-function $Q_+$ or $Q_-$. The Q-functions are polynomials of degree $2m_\pm$ parametrised by the Bethe roots $z^\pm_i$. Here $m_\pm$ is the magnon number and denotes the number of up (down) spins in the sea of down (up) spins. The Q-functions can be written as
\begin{equation}\label{eq:qfun}
 Q_\pm(z)=\frac{1}{2m_\pm-\LL \mp p\mp q}\,\prod_{i=1}^{m_\pm}(z-z^\pm_i)(z+z^\pm_i+1)\,.
\end{equation}  
We note that commonly the Q-functions are defined without the $z$-independent normalisation which drops from the TQ-equation. However, the Q-operator construction presented in the next section naturally incorporates the prefactor above. 
The Bethe equations follow from Baxter's TQ-equation. They take the form
\begin{equation}\label{eq:bethe}
 \frac{(z_i^\pm \pm p)(z_i^\pm \pm q)(z_i^\pm+1)^{2\LL}}{(z_i^\pm\mp p+1) (z_i^\pm\mp q+1)(z_i^\pm)^{2\LL}}=\prod_{\substack{k=1\\k\neq i}}^{m_\pm}\frac{(z_i^\pm-z_k^\pm+1)(z_i^\pm+z_k^\pm+2)}{(z_i^\pm-z_k^\pm-1)(z_i^\pm+z_k^\pm)}\,.
\end{equation} 
Finally, the nearest-neighbour Hamiltonian \eqref{eq:hamiltonian} is related to the transfer matrix via $\top'(0)=2pq\left(\LL+\mathbf{H}\right)$ and consequently belongs to the family of commuting operators. Here we identified $\xi=1/p$ and $\hat\xi=1/q$. Thus, the energy eigenvalues can be obtained in terms of the Q-functions from the derivative of $T$ at $z=0$.

\section{Q-operator construction}\label{sec:qop}
In this section we define the Q-operators $\qop_\pm$ of the open Heisenberg spin chain. We first introduce the Lax operators for the single-row monodromies and discuss some of their properties. Then we derive the boundary operators relevant to construct Q-operators from the boundary Yang-Baxter equation and finally define the Q-operators of the open Heisenberg spin chain. 

\subsection{Degenerate solutions to the Yang-Baxter equation}
As for the construction of the Q-operators for closed spin-$\frac{1}{2}$ Heisenberg spin chain \cite{Bazhanov2010a} we introduce the Lax operators
\begin{equation}\label{eq:plax}
 \plax_+(z)=\left(\begin{array}{cc}
              1&\oad\\
	      \oa&z+1+\NN
             \end{array}\right)\,,\quad\quad \plax_-(z)=\left(\begin{array}{cc}
              z+1+\NN&\oa\\
	      \oad&1
             \end{array}\right)\,.
\end{equation} 
Here the pair of oscillators $(\oa,\oad)$ obeys the usual commutation relations and $\NN$ is the number operator
\begin{equation}\label{eq:oscom}
 [\oa,\oad]=1\,,\qquad \NN=\oad\oa\,.
\end{equation} 
The Lax operators above are solutions of the Yang-Baxter equation
\begin{equation}\label{eq:rll}
 \rfund(x-y)(\plax_\pm(x)\otimes\mathbb{I})(\mathbb{I}\otimes \plax_\pm(y))=(\mathbb{I}\otimes \plax_\pm(y))(\plax_\pm(x)\otimes\mathbb{I}) \rfund(x-y)\,,
\end{equation} 
where $R$ is the fundamental R-matrix defined in \eqref{eq:rmatrix}. 
 In addition to the Lax operators $\plax_\pm$ in \eqref{eq:plax} we define 
 \begin{equation}
 \bar \plax_+(z)=\left(\begin{array}{cc}
              z-\NN&\oad\\
	      \oa&-1
             \end{array}\right)\,,\quad\quad \bar \plax_-(z)=\left(\begin{array}{cc}
              -1&\oa\\
	      \oad&z-\NN
             \end{array}\right)\,.
\end{equation} 
Together with the Lax operators in \eqref{eq:plax} they satisfy the unitarity and crossing-unitarity relations
\begin{equation}\label{eq:unit}
 \plax_\pm(z)\bar \plax_\pm(-z)=-z\,,\qquad \bar \plax_\pm^t(-z-2) \plax_\pm^t(z)=-z-1\,,
\end{equation} 
c.f.~\eqref{eq:unitR}. Here the superscript $t$ denotes the transposition of the $2\times2$ matrix which leaves the oscillators untouched.
Note that the Lax operators $\bar{\plax}_\pm$ can be related to $\plax_\pm$ by a particle-hole transformation and a subsequent multiplication of a diagonal matrix.
As the particle hole transformation does not change the commutation relations and the tensor product of any two diagonal matrices commutes with the R-matrix, we find that the Lax operators $\bar{\plax}_\pm$ satisfy the same Yang-Baxter equation as $\plax_\pm$, cf.~\eqref{eq:rll}. 

\subsection{Boundary Yang-Baxter equation}
In analogy to the transfer matrix construction we define the boundary operators
\begin{equation}\label{eq:bop}
 \KQL_\pm(z)=\frac{\Gamma(\mp p-z-1-\NN)}{\rho_\pm(z)} \,,\quad\quad  \KQR_\pm(z)=\frac{\hat{\rho}_\pm(z)}{\Gamma(\pm q-z-\NN)}\,,
\end{equation} 
which act diagonally in the oscillator space. Here, as discussed below, the normalisation reads
\begin{equation}\label{eq:knorm}
 \rho_\pm(z)=\Gamma(\mp p-z)\,,\quad\quad \hat{\rho}_\pm(z)=\Gamma(\pm q-z)\,.
\end{equation} 
We did not find these boundary operators in the literature.
The operators $\KQL_\pm$ and $\KQR_\pm$ are the counterparts of the  boundary matrices in \eqref{eq:kmatrices} and can be determined through the boundary Yang-Baxter equations 
\begin{equation}\label{eq:byberight}
\plax_\pm(x-y)\KQR_\pm(x)\plax_\pm(x+y) \KR (y)= \KR (y)\plax_\pm(x+y)\KQR_\pm(x)\plax_\pm(x-y)\,,
\end{equation} 
and
\begin{equation}\label{eq:bybeleft}
\bar \plax_\pm(y-x)\KQL_\pm(x)\bar \plax_\pm(-x-y-2) \KL (y)= \KL (y)\bar \plax_\pm(-x-y-2)\KQL_\pm(x)\bar \plax_\pm(y-x) \,.
\end{equation} 
Using the explicit expressions for boundary matrices $\KL$ and $\KR$ as well as the Lax operators $\plax_\pm$ and $\bar{\plax}_\pm$ we find that this equation is trivially satisfied for the diagonal entries in the $2\times 2$ space. Here we additionally assumed that the boundary operators $\KQL_\pm$ and $\KQR_\pm$ are diagonal, i.e.~only depend on the number operator $\NN=\oad\oa$. 
The remaining two defining relations arising from \eqref{eq:byberight} are conjugate to each other. The same holds true for the defining relations obtained from \eqref{eq:bybeleft}. They read
\begin{equation}
\oa\KQR_\pm(x)=(\pm q-x-1-\NN) \KQR_\pm(x) \oa\quad\text{and} \quad \KQL_\pm(x)\oa=\oa\KQL_\pm(x)(\mp p-x-1-\NN)\,.
\end{equation}
These equations are naturally solved by the Gamma function as given in \eqref{eq:bop}. 
The functions $\rho_\pm$ and $\hat\rho_\pm$ cannot be determined from the boundary Yang-Baxter equations \eqref{eq:byberight} and \eqref{eq:bybeleft}. The normalisation in \eqref{eq:knorm}, which renders the Q-operators to be polynomials in the spectral parameter, is obtained in Section~\ref{sec:traceoveraux}.

\subsection{Definition of the Q-operators}
We will now define the Q-operators. First, we introduce the double-row monodromies $\qh_\pm$
\begin{equation}\label{eq:drq}
\qh_\pm(z)=\qmon_\pm(z)\KQR_\pm (z)\hat\qmon_\pm(z)\,.
\end{equation} 
Here the boundary operators $\KQR_\pm$ were defined in \eqref{eq:bop}. The monodromies $\qmon_\pm$ and $\hat\qmon_\pm$ are built in analogy to \eqref{eq:mon} from the tensor product of the Lax operators $\plax^{[i]}$ in  \eqref{eq:plax} acting on the $i^\text{th}$ site of the spin chain and the ordinary product in the oscillator space as 
\begin{equation}\label{eq:qmons}
 \qmon_\pm(z)=\plax_\pm^{[1]}(z)\plax_\pm^{[2]}(z)\cdots \plax_\pm^{[\LL]}(z)\,,\qquad  \hat\qmon_\pm(z)=\plax_\pm^{[\LL]}(z)\plax_\pm^{[\LL-1]}(z)\cdots \plax_\pm^{[1]}(z)\,.
\end{equation} 
The double-row monodromy matrix \eqref{eq:drq} and the boundary operator $\KQL_\pm$ in \eqref{eq:bop} allow us to define the Q-operators $ \qop_\pm$ as 
\begin{equation}\label{eq:qop}
 \qop_\pm(z)=\tr \KQL_\pm (z)\qh_\pm(z)\,.
\end{equation} 
Here the trace is taken in the oscillator space. 
For convenience we give the explicit form of the Q-operators in \eqref{eq:qop}. They read
\begin{equation}
\begin{split}
 \qop_+(z)=\tr\, &\frac{\Gamma(-\NN-p-z-1)}{\Gamma(-p-z)} \left(\begin{array}{cc}
              1&\oad\\
	      \oa&z+1+\NN
             \end{array}\right)_1\cdots \left(\begin{array}{cc}
              1&\oad\\
	      \oa&z+1+\NN
             \end{array}\right)_\LL\\
              &\,\,\,\times  \frac{\Gamma(q-z)}{\Gamma(-\NN+q-z)} \left(\begin{array}{cc}
            1&\oad\\
	      \oa&z+1+\NN
             \end{array}\right)_\LL\cdots\left(\begin{array}{cc}
             1&\oad\\
	      \oa&z+1+\NN
             \end{array}\right)_1\,,
\end{split}
\end{equation} 
and
\begin{equation}
\begin{split}
 \qop_-(z)=\tr\, & \frac{\Gamma(-\NN+\bl-z-1)}{\Gamma(p-z)} \left(\begin{array}{cc}
              z+1+\NN&\oa\\
	      \oad&1
             \end{array}\right)_1\cdots \left(\begin{array}{cc}
               z+1+\NN&\oa\\
	      \oad&1
             \end{array}\right)_\LL\\
             &\,\,\,\times \frac{\Gamma(-q-z)}{\Gamma(-\NN-\br-z)} \left(\begin{array}{cc}
            z+1+\NN&\oa\\
	      \oad&1
             \end{array}\right)_\LL\cdots\left(\begin{array}{cc}
              z+1+\NN&\oa\\
	      \oad&1
             \end{array}\right)_1\,.
\end{split}
\end{equation} 
As shown in the next section, these operators are well-defined and in particular the trace yields a finite result. The entries of the Q-operators are polynomials in the spectral parameter arranged in blocks labelled by the magnon number. As also shown in the following, the Q-operators commute with the transfer matrix and satisfy Baxter's TQ-equation.

\section{Properties of the Q-operators}\label{sec:qprops}
In the following we discuss some properties of the Q-operators defined in Section~\ref{sec:qop}. First we show that the Q-operators commute with the transfer matrix and  discuss a symmetry among the Q-operators which is apparent in our construction.
Furthermore, we show that the trace over the oscillator space is well-defined and argue that the entries of the Q-operators are arranged in $L+1$ blocks corresponding to the magnon sectors $m_\pm$. 
Finally, we study the behaviour of the Q-operators when the spectral parameter $z$ is large. We find that each block is of maximal degree $z^{2m_\pm}$ in the spectral parameter and determine the overall prefactor apparent in our construction, cf.~\eqref{eq:qfun}.

\subsection{Commutativity}\label{sec:com}
\begin{figure}[hb]
\centering
 \includegraphics[width=\textwidth]{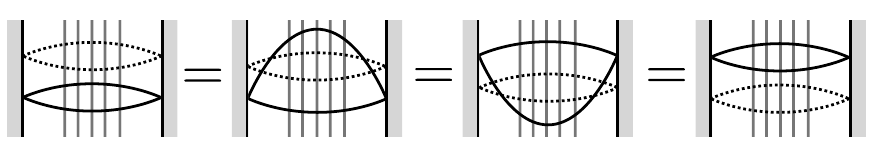}
 \caption{Schematic proof of the commutativity of $\top$ and $\qop_\pm$.}
 \label{fig:tq}
\end{figure}
In this section we show that the Q-operators in \eqref{eq:qop} commute with the transfer matrix  \eqref{eq:trans}. In particular, this is a consequence of the boundary Yang-Baxter equation for the double-row monodromies \eqref{bybef3} and 
\begin{equation}\label{eq:bybemon}
 \plax_\pm(x-y)\qh_\pm(x)\plax_\pm(x+y)\mathcal{U}(y)=\mathcal{U}(y)\plax_\pm(x+y)\qh_\pm(x)\plax_\pm(x-y)\,,
\end{equation} 
as defined in \eqref{eq:trans} and \eqref{eq:drq} respectively. The boundary Yang-Baxter equation \eqref{eq:bybemon} yields the fundamental commutation relations among the double-row monodromies of the Q-operators and the transfer matrix. It follows from the boundary Yang-Baxter equation in \eqref{eq:byberight} and the Yang-Baxter relations for the single-row monodromies
\begin{equation}\label{eq:rtt2}
\mathcal{M}(y)\qmon_\pm(x)\plax_\pm(x-y)=\plax_\pm(x-y)\qmon_\pm(x)\mathcal{M}(y) \,,
\end{equation} 
and
\begin{equation}\label{eq:rtt3}
\hat{\mathcal{M}}(y)\hat{\qmon}_\pm(x)\plax_\pm(x-y)= \plax_\pm(x-y)\hat{\qmon}_\pm(x)\hat{\mathcal{M}}(y)\,.
\end{equation} 
As a consequence of the unitarity relation for the R-matrix \eqref{eq:unitR} which yields $\hat{\mathcal{M}}(z)=(1-z^2)^\LL \mathcal{M}^{-1}(-z)$ the {\small RTT}-relations \eqref{eq:rtt2} and \eqref{eq:rtt3} imply that
\begin{equation}\label{eq:rtt1}
\hat{\qmon}_\pm(x)\plax_\pm(x+y)\mathcal{M}(y)=\mathcal{M}(y)\plax_\pm(x+y)\hat{\qmon}_\pm(x)\,,
\end{equation} 
and
\begin{equation}\label{eq:rtt4}
 \qmon_\pm(x)\plax_\pm(x+y)\hat{\mathcal{M}}(y)=\hat{\mathcal{M}}(y)\plax_\pm(x+y)\qmon_\pm(x)\,.
\end{equation} 

Now, it is  straightforward to show that the Q-operators commute with the transfer matrix. Using the explicit form of the transfer matrix and the Q-operators we find
\begin{equation}\label{eq:tqcom}
\begin{split}
\qop_\pm(x) \mathbf{T}(y)&= \trr\tr\big[  \qh_\pm(x)\mathcal{U}(y)\KL (y)\KQL_\pm (x)\big]\\
&=\frac{-1}{x+y+1} \trr\tr \big[\qh_\pm(x)\plax_\pm(x+y)\mathcal{U}(y)\KL(y)\bar\plax_\pm(-x-y-2)\KQL_\pm (x)\big]\\
&=\frac{1}{(x+y+1)(x-y)} \trr\tr\big[ \plax_\pm(x-y)\qh_\pm(x)\plax_\pm(x+y)\mathcal{U}(y)\\
&\qquad\qquad\qquad\qquad\qquad\quad\quad\times\KL (y)\bar\plax_\pm(-x-y-2)\KQL_\pm (x)\bar\plax_\pm(y-x)\big]\\
&=\frac{1}{(x+y+1)(x-y)} \trr\tr\big[ \mathcal{U}(y)\plax_\pm(x+y)\qh_\pm(x)\plax_\pm(x-y)\\
&\qquad\qquad\qquad\qquad\qquad\quad\quad\times\bar\plax_\pm(y-x)\KQL_\pm (x)\bar\plax_\pm(-x-y-2)\KL_a (y)\big]\\
&=\frac{-1}{x+y+1} \trr\tr\big[ \mathcal{U}(y)\plax_\pm(x+y)\qh_\pm(x)\KQL_\pm (x)\bar\plax_\pm(-x-y-2)\KL (y)\big]\\
&=\trr\tr\big[ \mathcal{U}(y)\qh_\pm(x)\KQL_\pm (x)\KL (y)\big]\\
&= \mathbf{T}(y)\qop_\pm(x)\,.
\end{split}
\end{equation} 
Here, we inserted the identity operator in the first and second step via the unitarity and crossing unitarity relation \eqref{eq:unit} using the cyclicity of the traces. Employing the boundary Yang-Baxter equation for the double-row monodromies \eqref{eq:bybemon}, the boundary Yang-Baxter equation in \eqref{eq:bybeleft} and \eqref{bybef2} for $\KQL_\pm$ and $\KL$ and subsequently using again the unitarity and crossing unitarity relation \eqref{eq:unit} yields the desired commutation relations
\begin{equation}
 [\mathbf{T}(x),\qop_\pm(y)]=0\,.
\end{equation} 
The manipulations explained above are schematically depicted in Figure~\ref{fig:tq}. Here each two crossing lines denote an R-matrix. 
 The gray vertical lines denote the quantum space while the dashed and the black line denote the auxiliary space of $\top$ and $\qop$ respectively with the corresponding K-matrices at the boundaries which are indicated by the black vertical lines to the left and right. The first and last equality make use of the unitarity and crossing unitarity relations as well as the {\small RTT}-relations above while the middle equation uses the boundary Yang-Baxter equations.

\subsection{Spin-flip symmetry}\label{sec:sym}
For the Heisenberg spin chain the two Q-operators $\qop_\pm$ can be related to each other by a similarity transformation in the quantum space and a change of sign in the boundary parameters $p$ and $q$. This follows from a relation among the Lax operators $\plax_\pm$ in  \eqref{eq:plax}  and the explicit form of the boundary operators in \eqref{eq:bop}. The Lax operators are related via 
\begin{equation}
 \left(\begin{array}{cc}
              0&1\\
	     1&0
             \end{array}\right)\plax_\pm(z)\left(\begin{array}{cc}
              0&1\\
	     1&0
             \end{array}\right)=\plax_\mp(z)\,.
\end{equation} 
Obviously, a similar relation holds for the single-row monodromies $\qmon_\pm$ and $\hat{\qmon}_\pm$ of the Q-operators involving the similarity transformation
\begin{equation}
 \mathcal{S}=\mathcal{S}^{-1}=\left(\begin{array}{cc}
              0&1\\
	     1&0
             \end{array}\right)\otimes\ldots\otimes\left(\begin{array}{cc}
              0&1\\
	     1&0
             \end{array}\right)\,.
\end{equation} 
As a consequence of the symmetry among the single-row monodromies and the form of the boundary operators $\KQR_\pm$ and $\KQL_\pm$ in \eqref{eq:bop} one immediately obtains a relation among the Q-operators $\qop_\pm$. It reads 
\begin{equation}\label{eq:qsym}
 \mathcal{S}\qop_\pm(z)\mathcal{S}^{-1}=\qop_\mp(z)\big|_{\substack{p\rightarrow -p\\q\rightarrow -q}}\,.
\end{equation} 
Finally, we note that the transfer matrix $\top$ is invariant under the similarity transformation and a subsequent change of sign in the boundary parameters $p$ and $q$.

In what follows we often focus on the Q-operator $\qop_+$ as it can be related to the $\qop_-$ using the relation in \eqref{eq:qsym}. 

\subsection{Block structure}\label{sec:block}
The Hamiltonian as well as the transfer matrix of the open Heisenberg spin chain are block diagonal, i.e. they do not contain entries that mix between different magnon numbers. This can be seen from the expansion of the boundary Yang-Baxter equation \eqref{bybef3} at order  $x^{2\LL+2}$ after building the transfer matrix from $\mathcal{U}(y)$, see \cite{Faddeev2007} where the calculation is done for the closed chain. We expect that also the Q-operators do not mix between states of different magnon numbers such that $[\qop_\pm (z),\sum_{i=1}^\LL\sigma^3_i]=0$. This is indeed true. It can be verified by evaluating the boundary Yang-Baxter equation \eqref{eq:bybemon} at order $y^{2L+2}$ and subsequently tracing over the oscillator space after multiplying by $\KQL_\pm$. 
However, instead of using the fundamental commutation relations that arise from \eqref{eq:bybemon} we demonstrate this property by evaluating the corresponding matrix elements of the Q-operators explicitly. This has the advantage of introducing some notation that will be needed in Section~\ref{sec:largez}.
From the definition of the Q-operators in \eqref{eq:qop} we find that the matrix elements read
\begin{equation}\label{eq:qopcomp}
 \big(\qop_\pm(z)\big)^{b_1b_2\ldots b_\LL}_{a_1a_2\ldots a_\LL}=\sum_{\{c\}=1}^2\tr \left[\KQL_\pm (z)\big(\qmon_\pm(z)\big)^{c_1c_2\ldots c_\LL}_{a_1a_2\ldots a_\LL}\KQR_\pm (z)\big(\hat \qmon_\pm(z)\big)_{c_1 c_{2}\ldots c_\LL}^{b_1 b_{2}\ldots b_\LL}\right]\,, 
\end{equation} 
where
\begin{equation}\label{eq:qqop}
\qop_\pm(z)=\sum_{\{a\},\{b\}=1}^2  \big(\qop_\pm(z)\big)^{b_1b_2\ldots b_\LL}_{a_1a_2\ldots a_\LL}\, e_{a_1b_1}\otimes e_{a_2b_2}\otimes \ldots\otimes e_{a_\LL b_\LL}\,.
\end{equation} 
Here we defined the components of the single-row monodromies $\qmon_\pm$ and $\hat \qmon_\pm$ in analogy to \eqref{eq:qqop}. They  are given in terms of the components of the Lax operators $\plax_\pm(z)=\sum_{a,b=1}^2 \big(\plax_\pm(z)\big)_{ab}\, e_{ab}$ in the quantum space as 
\begin{equation}\label{eq:moncom}
  \big(\qmon_\pm(z)\big)^{b_1b_2\ldots b_\LL}_{a_1a_2\ldots a_\LL}= \big(\hat\qmon_\pm(z)\big)_{a_\LL a_{\LL-1}\ldots a_1}^{b_\LL b_{\LL-1}\ldots b_1}=\big(\plax_\pm(z)\big)_{a_1b_1}\big(\plax_\pm(z)\big)_{a_2b_2}\cdots \big(\plax_\pm(z)\big)_{a_\LL b_\LL}\,.
\end{equation} 

In order to show that the Q-operators are arranged in blocks as well as for the discussion in Section~\ref{sec:largez} it is convenient to introduce four sets
\begin{equation}
\set_{ij}^{(a,b)}  =  \left\{ l\in\left\{ 1,2,\dots,\LL\right\} |a_{l}=i\,\wedge\,b_{l}=j\right\}\,,  
\end{equation}
of cardinality $|\set_{ij}^{(a,b)}|$ and $i,j\in\{1,2\}$. These four sets are in one-to-one correspondence to the indices of the quantum space, cf.~\eqref{eq:qqop}, and thus uniquely label a given matrix element. From \eqref{eq:moncom} and the explicit form of the Lax operators \eqref{eq:plax} we know that the monodromy $\big(\qmon_+(z)\big)^{c_1c_2\ldots c_\LL}_{a_1a_2\ldots a_\LL}$ is a sequence of $|\set_{12}^{(a,c)}|$ operators $\oad$, $|\set_{21}^{(a,c)}|$ operators $\oa$ and $|\set_{22}^{(a,c)}|$  operators $\left(z+1+\NN\right)$. Similarly, $\big(\hat \qmon_\pm(z)\big)_{c_1 c_{2}\ldots c_\LL}^{b_1 b_{2}\ldots b_\LL}$ is a sequence of $|\set_{12}^{(c,b)}|$ operators $\oad$, $|\set_{21}^{(c,b)}|$ operators $\oa$ and $|\set_{22}^{(c,b)}|$  operators $\left(z+1+\NN\right)$. As the boundary operators $\KQL_+$ and $\KQR_+$ only depend on the number operator $\NN$ we find that the trace in \eqref{eq:qopcomp} is non-vanishing if 
\begin{equation}\label{eq:rel1}
|\set_{12}^{(a,c)}|+|\set_{12}^{(c,b)}|= |\set_{21}^{(a,c)}|+|\set_{21}^{(c,b)}|\,,
\end{equation} 
i.e. the total number of creation operators coincides with the total number of annihilation operators in $\big( \qmon_\pm(z)\big)_{c_1 c_{2}\ldots c_\LL}^{b_1 b_{2}\ldots b_\LL}$ and $\big(\hat \qmon_\pm(z)\big)_{c_1 c_{2}\ldots c_\LL}^{b_1 b_{2}\ldots b_\LL}$. The cardinality of the sets can be related to the magnon numbers of the in and out state, i.e. the number of indices with value ``$2$'' in  $(a_1,a_2,\ldots,a_\LL)$ and $(b_1,b_2,\ldots,b_\LL)$ respectively, via 
\begin{equation}\label{eq:rel2}
 m_+^a=|\set_{21}^{(a,c)}|+|\set_{22}^{(a,c)}|\,,\qquad  m_+^b=|\set_{12}^{(c,b)}|+|\set_{22}^{(c,b)}|\,.
\end{equation} 
After substituting \eqref{eq:rel2} in \eqref{eq:rel1} and using the identity $|\set_{12}^{(a,c)}|+|\set_{22}^{(a,c)}|= |\set_{21}^{(c,b)}|+|\set_{22}^{(c,b)}|$ we find that 
\begin{equation}\label{eq:rel3}
  m_+^a= m_+^b\equiv m_+\,.
\end{equation} 
A similar argument holds true for $\qop_-$. 

Thus, we find that the Q-operators $\qop_\pm$ are organised in $L+1$ blocks with $m_\pm=0,1,\ldots,\LL$ where the size of each block is $\binom{\LL}{m_\pm}\times \binom{\LL}{m_\pm}$. 

\subsection{Tracing over the auxiliary space}\label{sec:traceoveraux}
In this section we show that the trace over the non-vanishing entries of the Q-operators is well-defined and determine the normalisation $\rho_\pm$ and $\hat\rho_\pm$, cf.~\eqref{eq:knorm},  which renders all entries polynomials in the spectral parameter.

First, we note that the entries of the Q-operators can be expanded in terms of the oscillators. 
Thus, after normal ordering of a given sequence in the monodormy $\qmon_+$ and anti-normal ordering the sequence in the monodormy $\hat \qmon_+$ as well as using the linearity of the trace we find that 
by construction all entries that contribute to the Q-operators in \eqref{eq:qop} can be brought to the form
\begin{equation}\label{eq:tracedef}
\mathcal{P}^\pm_{r_1,r_2,s_1,s_2}= \tr\left[ \KQL_\pm(z)\,\oad^{r_1}\oa^{r_2}\, \KQR_\pm (z)\,\oa^{s_2}\oad^{s_1}\right]\,,
\end{equation} 
with $r_1,r_2,s_1,s_2\in \mathbb{N}^0$.
Here, as discussed above, the trace is non-vanishing if there are as many creation as annihilation operators
\begin{equation}
r_1+s_1=r_2+s_2\,,
\end{equation} 
cf.~\eqref{eq:rel1}.
After inserting the explicit form of the boundary operators \eqref{eq:bop} and reordering the oscillators using the commutation relations in \eqref{eq:oscom} the argument in the trace of \eqref{eq:tracedef}  can be written in terms of the number operator $\NN$ as
\begin{equation}\label{eq:tracedeff}
\mathcal{P}^\pm_{r_1,r_2,s_1,s_2}=\frac{\hat\rho_\pm(z)}{\rho_\pm(z)}\sum_{k=0}^{r_1+s_1}S_1(r_1+s_1,k) \tr\left[ \frac{\Gamma(s_1\mp p-z-1-\NN)}{\Gamma(s_2\pm q-z-\NN)}\,\NN^{k}\right]\,.
\end{equation} 
Here, we used the relation $\oad^j\oa^j=\sum_{k=0}^j S_1(j,k)\NN^k$ with $S_1$ the Stirling numbers of first kind \cite{AbramowitzStegun:1964}. The trace in \eqref{eq:tracedeff} can be evaluated using the relation 
\begin{equation}\label{eq:sumf}
\tr \,\frac{\Gamma(x_\pm-\NN)}{\Gamma(y_\pm-\NN)}\,\NN^k=\begin{cases}
                                                \frac{\Gamma(1+x_\pm)}{(1+x_\pm-y_\pm)\Gamma(y_\pm)}&k=0\\
                                                \\
                                                -\frac{\Gamma(x_\pm)}{\Gamma(y_\pm-x_\pm)}\sum_{j=0}^k(-1)^j\, j!\,S_2(k+1,j+1)\,\frac{\Gamma(y_\pm-x_\pm-1-j)}{\Gamma(y_\pm-1-j)}&k\in\mathbb{N}^+
                                               \end{cases}\,,
\end{equation} 
where $S_2$ denotes the Stirling numbers of second kind \cite{AbramowitzStegun:1964} and we used the abbreviations
\begin{equation}
 x_\pm=s_1\mp p-z-1\,,\qquad y_\pm=s_2\pm q-z\,.
\end{equation} 
The relation in \eqref{eq:sumf} can be obtained by rewriting the  trace as a hypergeometric function which can be related to a polylogarithm using Euler's integral transform. The series expansion of the polylogarithm of non-positive integer order and the integral representation of the Beta-function yields \eqref{eq:sumf}. The derivation can be found in Appendix~\ref{app:trace}. From \eqref{eq:sumf} it is straightforward to evaluate the sum in \eqref{eq:tracedeff} using the identity
\begin{equation}
 \sum_{i=j}^{r_1+s_1}S_1(r_1+s_1,i)S_2(i+1,j+1)=(j+1)\delta_{r_1+s_1,j+1}+\delta_{r_1+s_1,j}\,.
\end{equation} 
It can be shown using the recurrence relation for the Stirling numbers of second kind and their inverse relation \cite{AbramowitzStegun:1964}. Finally, we find
\begin{equation}\label{eq:tracepol}
\begin{split}
 \mathcal{P}^\pm_{r_1,r_2,s_1,s_2}&=(r_2+s_2)!\,\frac{\Gamma(\mp p\mp q+s_1-s_2)}{\Gamma(\mp p\mp q+s_1+r_2+1)}\frac{\hat\rho_\pm(z,q)}{\rho_\pm(z,p)}\frac{\Gamma(\mp p-z+s_1)}{\Gamma(\pm q-z-r_2)}\,.
 \end{split}
\end{equation} 
Obviously, as $s_1\geq0$ and $r_2\geq0$, the normalisation in \eqref{eq:knorm} renders $ \mathcal{P}^\pm$ a polynomial in $z$ of degree $s_1+r_2$. Therefore, also all entries of $\qop_\pm$ are polynomials in the spectral parameter $z$. In the next section we will determine an upper bound for the degree of the polynomials in a given magnon sector.

\subsection{Leading-order $z$ behaviour}\label{sec:largez}
In this section we show that the Q-operators are polynomials of maximal degree $z^{2 m_\pm}$ in the corresponding magnon sectors and determine the $m_\pm$-dependent prefactor of the Q-functions, cf.~\eqref{eq:qfun}. 

In order to determine the leading order $z$ behaviour we focus on a given intermediate state $(c_1,c_2,\ldots,c_\LL)$ in the sum that yields the components of the Q-operator $\qop_+$. We define 
\begin{equation}
  \cc(z;a,b,c)=\tr \left[\KQL_+ (z)\big(\qmon_+(z)\big)^{c_1c_2\ldots c_\LL}_{a_1a_2\ldots a_\LL}\KQR_+ (z)\big(\hat \qmon_+(z)\big)_{c_1 c_{2}\ldots c_\LL}^{b_1 b_{2}\ldots b_\LL}\right]\,,
\end{equation} 
cf.~\eqref{eq:qopcomp}.
Since we know the number of $\left(z+1+\NN\right)$ in the sequence of a given matrix element of the single-row monodromies $\qmon_+(z)$ and $\hat{\qmon}_+(z)$, cf.~Section~\ref{sec:block}, we can expand them in powers of $(z+1)$. The expansion for the components of $\qmon_+(z)$ reads
\begin{equation}\label{eq:interme}
\big(\qmon_+(z)\big)^{c_1c_2\ldots c_\LL}_{a_1a_2\ldots a_\LL}= \sum_{k=0}^{|\set_{22}^{(a,c)}|} \left(z+1\right)^{|\set_{22}^{(a,c)}|-k} \big(\qmon_+^{(k)}\big)^{c_1c_2\ldots c_\LL}_{a_1a_2\ldots a_\LL}\,.
\end{equation}
Here $\big(\qmon_+^{(k)}\big)^{c_1c_2\ldots c_\LL}_{a_1a_2\ldots a_\LL}$ is a sum of $\binom{|I_{22}^{(a,c)}|}{k}$ sequences that only contain oscillators. 
Moreover, each sequence contains the same numbers of $\oa$ and $\oad$ as $\big(\qmon_+(z)\big)^{c_1c_2\ldots c_\LL}_{a_1a_2\ldots a_\LL}$, cf.~Section~\ref{sec:block}, and $k$ operators $\NN$.
A similar expansion holds true for the components of the monodromy $\hat \qmon_+$. It follows that the contribution to the sum $\eqref{eq:interme}$ can be written as 
\begin{equation}\label{eq:qcomexp}
\begin{split}
 \cc(z;a,b,c)&=\sum_{k_1=0}^{|\set_{22}^{(a,c)}|} \sum_{k_2=0}^{|\set_{22}^{(c,b)}|} (z+1)^{|\set_{22}^{(a,c)}|+|\set_{22}^{(c,b)}|-k_1-k_2}\\
&\qquad\qquad\qquad\times\tr \left[\KQL_+ (z)\big(\qmon_+^{(k_1)}\big)^{c_1c_2\ldots c_\LL}_{a_1a_2\ldots a_\LL}\KQR_+ (z)\big(\hat \qmon_+^{(k_2)}\big)_{c_1 c_{2}\ldots c_\LL}^{b_1 b_{2}\ldots b_\LL}\right]\,.
\end{split}
\end{equation}

When evaluating the trace using \eqref{eq:tracepol}, the leading order in $z$ of $\cc$ can be extracted by normal ordering the sequences in $\big(\qmon_+^{(k_1)}\big)^{c_1c_2\ldots c_\LL}_{a_1a_2\ldots a_\LL}$ and picking the terms with the highest power of $\oa$, while anti-normal ordering $\big(\hat \qmon_+^{(k_2)}\big)_{c_1 c_2 \ldots c_\LL}^{b_1 b_2 \ldots b_\LL}$ and picking the highest power of $\oad$. This is equivalent to neglecting all the quantum corrections while normal (anti-normal) ordering which we denote by $:\big(\qmon_+^{(k_1)}\big)^{c_1c_2\ldots c_\LL}_{a_1a_2\ldots a_\LL} :$ and $\,\bar{:}\;\big(\hat \qmon_+^{(k_2)}\big)_{c_1 c_2 \ldots c_\LL}^{b_1 b_2 \ldots b_\LL}\; \bar{:}\,$ respectively.  As a consequence the large-$z$ contribution of a given intermediate state is 
\begin{equation}\label{eq:qcomexpordered}
\begin{split}
\cc (z,a,b,c) &\stackrel{z\gg1}{=} \sum_{k_1=0}^{|\set_{22}^{(a,c)}|} \sum_{k_2=0}^{|\set_{22}^{(c,b)}|} z^{|\set_{22}^{(a,c)}|+|\set_{22}^{(c,b)}|-k_1-k_2} \\
&\qquad\qquad\qquad\times\tr \left[\KQL_+ (z):\big(\qmon_+^{(k_1)}\big)^{c_1c_2\ldots c_\LL}_{a_1a_2\ldots a_\LL}:\KQR_+ (z)\; \bar{:}\;\big(\hat \qmon_+^{(k_2)}\big)_{c_1 c_2 \ldots c_\LL}^{b_1 b_2 \ldots b_\LL}\; \bar{:}\;\right]\,,
\end{split}
\end{equation}
where 
\begin{equation}
\begin{split}
:\big(\qmon_+^{(k_1)}\big)^{c_1c_2\ldots c_\LL}_{a_1a_2\ldots a_\LL} :&
=  \binom{|\set_{22}^{(a,c)}|}{k_1}\,  \oad^{|\set_{12}^{(a,c)}|+k_1}\, \oa^{|\set_{21}^{(a,c)}|+k_1}\,,
\end{split}
\end{equation}
and
\begin{equation}
\bar{:}\;\big(\hat \qmon_+^{(k_2)}\big)_{c_1 c_2 \ldots c_\LL}^{b_1 b_2 \ldots b_\LL}\; \bar{:}\;=
\binom{|\set_{22}^{(c,b)}|}{k_2}\,\oa^{|\set_{21}^{(c,b)}|+k_2}\, \oad^{|\set_{12}^{(c,b)}|+k_2}\;.
\end{equation} 
Using the trace formula in \eqref{eq:tracepol} we can evaluate the right-hand-side of \eqref{eq:qcomexpordered}. After neglecting further subleading contributions that arise from the trace formula we evaluate the sums for $k_1$ and $k_2$ using the relation \eqref{eq:GammaSumTrick1} and find
\begin{equation}\label{eq:largezQfixedc}
\begin{split}
\cc_{\loz} (z,|\set_{ij}^{(a,c)}|,|\set_{ij}^{(c,b)}|) &= z^{|\set_{21}^{(a,c)}|+|\set_{22}^{(a,c)}|+|\set_{12}^{(c,b)}|+|\set_{22}^{(c,b)}|}\,(-1)^{|\set_{12}^{(c,b)}|+|\set_{21}^{(a,c)}|}\,\Gamma(|\set_{21}^{(a,c)}|+|\set_{21}^{(c,b)}|+1)  \\
&\quad\times \frac{\Gamma(- p- q+|\set_{22}^{(a,c)}|+|\set_{22}^{(c,b)}|+|\set_{12}^{(c,b)}|-|\set_{21}^{(c,b)}|)}{\Gamma(- p- q+|\set_{21}^{(a,c)}|+|\set_{22}^{(a,c)}|+|\set_{12}^{(c,b)}|+|\set_{22}^{(c,b)}|+1)} \, .
\end{split} 
\end{equation}
After substituting the relation \eqref{eq:rel2} we find that the order of  $z$ in \eqref{eq:largezQfixedc} and consequently of a given magnon block of $\qop_+$  is $2m_+=|\set_{21}^{(a,c)}|+|\set_{22}^{(a,c)}|+|\set_{12}^{(c,b)}|+|\set_{22}^{(c,b)}|$. An analogous derivation of the block structure holds for $\qop_-$ but it also follows from the spin-flip symmetry discussed in Section~\ref{sec:sym}. Hence, in every magnon block $m_\pm$ of the Q-operators $\qop_\pm$ the maximal degree of the polynomial entries is $z^{2m_\pm}$.

Last but not least, we would like to determine the $z$-independent prefactor of the Q-functions in \eqref{eq:qfun}. To obtain it we sum over the intermediate states $(c_1, c_2, \ldots c_\LL)$ of $\mathcal{Y}_{\loz}$ which yields the leading order in $z$ of the Q-operator $\qop_+$
\begin{equation}
  \big(\qop_+(z)\big)^{b_1b_2\ldots b_\LL}_{a_1a_2\ldots a_\LL} \stackrel{z\gg1}{=}  \sum_{\{c\}=1}^2 \cc_{\loz} (z, |\set_{ij}^{(a,c)}|,|\set_{ij}^{(c,b)}|)\,.
\end{equation} 
Here the sum over $\{c\}$ denotes the sum over all indices $c_1,c_2,\ldots,c_\LL$.
Let us introduce the notation $n_{ij}=|\set_{ij}^{(a,b)}|$ and define $c_{ij}=|\set_{ij}^{(a,b)}\cap(\set_{12}^{(a,c)}\cup\set_{22}^{(a,c)})|$, where $\set_{12}^{(a,c)}\cup\set_{22}^{(a,c)}\subseteq \{1,2,\ldots,L\}$ simply denotes the indices whose value for the intermediate state $(c_1,c_2,\ldots,c_\LL)$ is ``$2$''. All cardinalities $|\set_{ij}^{(a,c)}|$ and $|\set_{ij}^{(c,b)}|$ can be expressed in terms of $n_{ij}$ and $c_{ij}$ using the relations
\begin{equation}
 \begin{array}{cc}
 |\set_{i1}^{(a,c)}|=n_{i1}+n_{i2}-c_{i1}-c_{i2}\,,&\qquad  |\set_{i2}^{(a,c)}|=c_{i1}+c_{i2}\,,\\ 
 &\\
  |\set_{1i}^{(c,b)}|=n_{1i}+n_{2i}-c_{1i}-c_{2i}\,,&\qquad  |\set_{2i}^{(c,b)}|=c_{1i}+c_{2i}\,.\\ 
 \end{array}
\end{equation} 
We note that $\mathcal{Y}_{\loz}$ in \eqref{eq:largezQfixedc} only depends on the cardinalities $|\set_{ij}^{(a,c)}|$ and $|\set_{ij}^{(c,b)}|$ and not on the actual ordering of the components in the single-row monodromies. Thus we can parametrise the sum using the variables $c_{ij}$. The sum over the intermediate states takes the form
\begin{equation}\label{eq:cvars}
  \big(\qop_+(z)\big)^{b_1b_2\ldots b_\LL}_{a_1a_2\ldots a_\LL} \stackrel{z\gg1}{=}\sum_{c_{11}=0}^{n_{11}}\binom{n_{11}}{c_{11}}\sum_{c_{12}=0}^{n_{12}}\binom{n_{12}}{c_{12}}\sum_{c_{21}=0}^{n_{21}}\binom{n_{21}}{c_{11}}\sum_{c_{22}=0}^{n_{22}}\binom{n_{22}}{c_{22}} \cc_{\loz} (z, n_{ij},c_{ij})\,.
\end{equation}
From the explicit form of $\mathcal{Y}_{\loz}$ in terms of the variables introduced in \eqref{eq:cvars} we find that the alternating sums of the binomials over $c_{12}$ and $c_{21}$ vanish if $n_{12}\neq 0$ and $n_{21}\neq0$. This is the case for any off-diagonal matrix element of the Q-operator. The two remaining sums over $c_{11}$ and $c_{22}$ can be evaluated using the relation \eqref{eq:GammaSumTrick2}. We find
\begin{equation}\label{eq:prefactor}
 \big(\qop_\pm(z)\big)^{b_1b_2\ldots b_\LL}_{a_1a_2\ldots a_\LL} \stackrel{z\gg1}{=}  \frac{z^{2m_\pm}}{2m_\pm-\LL\mp p\mp q}\,\delta_{ a_1 , b_1}\delta_{ a_2 , b_2}\cdots\delta_{ a_\LL , b_\LL}\, .
\end{equation}
Here the leading order $z$ behaviour of the Q-operator $\qop_-$ can either be obtained from a similar calculation or by applying the relation among the Q-operators discussed in Section~\ref{sec:sym}. 

As a consequence of the results obtained in this section the eigenvalues of the Q-operators $\qop_\pm$ are polynomials of degree $m_\pm$ in the spectral parameter $z$ with the leading coefficient given in \eqref{eq:prefactor}.  

\section{From Q-operators to Q-functions}\label{sec:tq}
In this section we prove that the Q-operators satisfy Baxter's TQ-equation on the operatorial level and discuss that as a consequence the roots of the eigenvalues of the Q-operators satisfy the Bethe equation. 

In order to show that the operators $\top$ and $\qop_\pm$ satisfy the TQ-equation, cf.~\eqref{eq:baxtereqn}, we take inspiration from Baxter's original method applied to construct the Q-operator of the eight-vertex model \cite{Baxter2007}, see also \cite{Pasquier1992,Sklyanin2007} where a similar method has been used.
 The idea is to multiply $\top$ and $\qop_\pm$ as given in \eqref{eq:trans} and \eqref{eq:qop} respectively and to bring the resulting monodromies and boundary matrices to an upper triangular form in the auxiliary space of the transfer matrix to finally take the trace. To do so we use the remarkable identity 
\begin{equation}\label{eq:lldec}
 \mathcal{L}^{[i]}_\a(z)\plax_+^{[i]}(z)=\mathbb{G}_{\a+}\,\mathbb{L}^{[i]}_{\a+}(z)\,
             \mathbb{G}^{-1}_{\a+}   \,,      
\end{equation} 
see also \cite{BoosCommun.Math.Phys.272:263-2812007}.
Here $\mathbb{L}^{[i]}_{\a+}$ is an upper triangular matrix in the auxiliary space of the transfer matrix denoted by ``$\a$'' with entries in the auxiliary space of the Q-operator ``$+$'' and the spin chain site $i$. It reads
\begin{equation}\label{eq:ltri}
 \mathbb{L}^{[i]}_{\a+}(z)=\left(\begin{array}{cc}
              (z+1)\plax_+^{[i]}(z-1)&e^{[i]}_{21}+\oad e^{[i]}_{22}\\
	     0&z\,\plax_+^{[i]}(z+1)
             \end{array}\right)_{\a}\,.
\end{equation} 
Furthermore, $\mathbb{G}_{\a+}$ is a lower triangular matrix in the auxiliary space of the transfer matrix with entries in the auxiliary space of the Q-operator 
\begin{equation}\label{eq:G}
\mathbb{G}_{\a+}=\left(\begin{array}{cc}
              1&0\\
	     \oa&1
             \end{array}\right)_{\a}\,.
\end{equation} 
A direct computation shows that similar formulae exist for the combinations of boundary operators
\begin{equation}\label{eq:KRdec}
 \KR_\a (z) \plax_+^{[\a]}(2z)\KQR_+(z)=\mathbb{G}_{\a+}\,\hat{\mathbb{K}}_{\a+}(z)\,
             \mathbb{G}^{-1}_{\a+}\,,
\end{equation} 
where 
\begin{equation}\label{eq:KRtri}
 \hat{\mathbb{K}}_{\a+}(z)=\left(\begin{array}{cc}
              (q+z)\KQR_+(z-1)&(q+z)\,\oad \,\KQR_+(z)\\
	     0&2z(q-z-1)\KQR_+(z+1)
             \end{array}\right)_\a\,,
\end{equation} 
as well as
\begin{equation}\label{eq:KLdec}
  \KQL_+(z)\bar\plax_+^{[\a]}(-2z-2)\KL_\a (z)=\mathbb{G}_{\a+}\,\mathbb{K}_{\a+}(z)\,
             \mathbb{G}^{-1}_{\a+}\,,
\end{equation} 
where
\begin{equation}\label{eq:KLtri}
 \mathbb{K}_{\a+}(z)=\left(\begin{array}{cc}
              -2(p+z)(z+1)\KQL_+(z-1)&(p-z-1)\,\KQL_+(z)\,\oad \\
	     0&-(p-z-1)\KQL_+(z+1)
             \end{array}\right)_\a\,.
\end{equation} 
Using the relations above it is straightforward to show that the TQ-equation is satisfied by the operators $\top$ and $\qop_+$. From \eqref{eq:tqcom} we know that
\begin{equation}\label{eq:tqtri}
\begin{split}
\mathbf{T}(z)\qop_+(z)&=\frac{-1}{2z+1} \trr\tr\big[ \mathcal{U}_\a(z)\plax_+(2z)\qh_+(z)\KQL_+ (z)\bar\plax_+(-2z-2)\KL_\a (z)\big]\\
&=\frac{-1}{2z+1} \trr\tr\big[ \mathcal{M}_\a(z)\qmon_+(z)\KR_\a(z)\plax_+(2z)\KQR_+(z)\\
&\qquad\qquad\qquad\qquad\qquad\times\hat{\mathcal{M}}_\a(z)\hat{\qmon}_+(z)\KQL_+ (z)\bar\plax_+(-2z-2)\KL_\a (z)\big]\,,
\end{split}
\end{equation} 
where we used the explicit form of the double-row monodromies and the {\small RTT}-relation \eqref{eq:rtt4}.
Furthermore, we note that 
\begin{equation}
  \mathcal{M}_a(z)\qmon_+(z)=\mathbb{G}_{a+}\,\mathbb{L}^{[1]}_{a+}(z)\mathbb{L}^{[2]}_{a+}(z)\cdots \mathbb{L}^{[\LL]}_{a+}(z)\,
             \mathbb{G}^{-1}_{a+}  \,,
\end{equation} 
and
\begin{equation}
  \hat{\mathcal{M}}_a(z)\hat{\qmon}_+(z)=\mathbb{G}_{a+}\,\mathbb{L}^{[\LL]}_{a+}(z)\mathbb{L}^{[\LL-1]}_{a+}(z)\cdots \mathbb{L}^{[1]}_{a+}(z)\,
             \mathbb{G}^{-1}_{a+}  \,.
\end{equation} 
Finally, using the decompositions involving the boundary operators \eqref{eq:KRdec} and \eqref{eq:KLdec} we can evaluate the trace in the auxiliary space of the transfer matrix which is written as a product of only upper triangular matrices. We find 
\small
\begin{equation}\label{eq:tqproof}
\begin{split}
\mathbf{T}(z)\qop_+(z)&=\hspace{-2pt}\trr\hspace{-2pt}\left(\begin{array}{cc}
              (q+z)(p+z)\frac{(z+1)^{2\LL+1}}{z+\frac{1}{2}}\qop_+(z-1)&*\\
	     0&(q-z-1)(p-z-1)\frac{z^{\LL+1}}{z+\frac{1}{2}}\qop_+(z+1)
             \end{array}\right)_\a\\
&=(q+z)(p+z)\frac{(z+1)^{2\LL+1}}{z+\frac{1}{2}}\qop_+(z-1)+(q-z-1)(p-z-1)\frac{z^{\LL+1}}{z+\frac{1}{2}}\qop_+(z+1)\,.
\end{split}
\end{equation} 
\normalsize
Here $*$ denotes a non-zero entry of the upper triangular matrix. This is Baxter's TQ-equation on the operatorial level involving $\qop_+$.
Using the symmetry between $\qop_+$ and $\qop_-$ discussed in Section~\ref{sec:sym} and noting that 
\begin{equation}
 \mathcal{S}\top(z)\mathcal{S}^{-1}=\top(z)\big|_{\substack{p\rightarrow -p\\q\rightarrow -q}}\,,
\end{equation} 
we obtain the TQ-equation for $\qop_-$. Thus we derived the TQ-equation in \eqref{eq:baxtereqn} for operators
\small
\begin{equation}\label{eq:baxtereqn2}
\begin{split}
 \top(z)\qop_\pm(z)=(z\pm p)(z\pm q)\frac{(z+1)^{2\LL+1}}{z+\half}\qop_\pm(z-1)+(z\mp p+1) (z\mp q+1)\frac{z^{2\LL+1}}{z+\half}\qop_\pm(z+1)\,.
 \end{split}
\end{equation}
\normalsize
From Section~\ref{sec:largez} we know that the eigenvalues of the Q-operators are polynomials in the spectral parameter of degree $m_\pm$ with the leading coefficient given in \eqref{eq:prefactor}. Furthermore, we note that  on the level of the eigenvalues the TQ-equation implies that
\begin{equation}
\frac{Q_\pm(z)}{Q_\pm(z-1)}=\frac{Q_\pm(-z-1)}{Q_\pm(-z)}\,.
\end{equation} 
This is a consequence of the crossing symmetry of the transfer matrix \eqref{eq:tcros}. It follows that the fraction above is of the form $Q_\pm(z)/Q_\pm(z-1)=\prod_{i=1}^{2m_\pm}(z-z^\pm_i)/(z+z^\pm_i)$ with $z_{i+m_\pm}^\pm=-z_{i}^\pm-1$ for $i=1,2,\ldots,m_\pm$. Hence we obtain
\begin{equation}\label{eq:qfun2}
 Q_\pm(z)=\frac{1}{ 2m_\pm-\LL \mp p\mp q}\,\prod_{i=1}^{m_\pm}(z-z^\pm_i)(z+z^\pm_i+1)\,.
\end{equation}  
Finally, evaluating the TQ-equation at the roots of the polynomials $Q_\pm$ we find that the roots satisfy the Bethe equations \eqref{eq:bethe}. Thus we conclude that the eigenvalues of the Q-operators are given by Baxter Q-functions which are polynomials of degree $2m_\pm$ of the form \eqref{eq:qfun2} with zeros at the Bethe roots.

\section{Summary and outlook}\label{sec:conc}
In this article we defined Q-operators for the open Heisenberg spin chain with diagonal boundary conditions.

Our method combines Sklyanin's formalism for integrable open spin chains \cite{Sklyanin1988} and the Q-operator construction for the closed Heisenberg spin chain \cite{Bazhanov2010a,Bazhanov2010,Frassek2011}. 
An obstacle that immediately arises is that the construction of the transfer matrix as presented in \cite{Sklyanin1988} assumes certain symmetries of the R-matrix and most of them are absent in the Lax operators used for the Q-operators construction \cite{Bazhanov2010a}. Thus, a priori it is not clear which boundary Yang-Baxter equations and unitarity relations have to be satisfied in order to build members of the commuting family of operators from the two bulk Lax matrices used in \cite{Bazhanov2010a}. 
The boundary Yang-Baxter equations and unitarity relations relevant to construct such operators that commute with the transfer matrix of the open Heisenberg spin chain, cf.~Section~\ref{sec:com}, were presented in Section~\ref{sec:qop}. These relations along with the solutions to the boundary Yang-Baxter equations \eqref{eq:bop} led us to define two operators which commute with the transfer matrix.
To show that the constructed operators are well defined we studied the evaluation of the trace in the auxiliary space which is more involved than for the case of the closed chain. We found that the operators are of block structure and polynomials in the spectral parameter governed by the magnon number. Finally, we showed that the defined operators satisfy Baxter's TQ-equation on the operatorial level. This allowed us to deduce the form of their eigenvalues and identify them as Baxter Q-operators.

We found that the Q-operators constructed here naturally incorporate a $z$-independent prefactor in the corresponding Q-functions. It depends on the length, the magnon number and the boundary parameters $p$ and $q$. In principle, this prefactor can be absorbed by redefining the monodromies for the Q-operator. However, naively this would spoil the locality which is manifest in our definition. 

It would be interesting to study the boundary operators introduced here in the context of the twisted Yangian \cite{molevbook} and derive further underlying functional relations for the Q-operators, cf.~\cite{Bazhanov2010a,Bazhanov2010} for the case of the closed chain. As for closed chains it should be possible for the open chain to obtain the Q-operators from transfer matrices by taking certain limits in the auxiliary space, see e.g.~\cite{Boos2010,Khoroshkin2014} where the relevant Lax operators were obtained from the universal R-matrix \cite{Khoroshkin1991}. It may also be interesting to study how the Hamiltonian can be obtained from the Q-operators as studied for the closed chain in \cite{Frassek2013} as well as the relation to the constructions of Q-operators for the open ASEP in \cite{Lazarescu2014}, the derivation of the TQ-equation in \cite{Yang2006} and the integral operator approach as studied for the closed chain in \cite{Chicherin2011,Chicherin2012b}. 

We hope that the construction generalises to the case of non-diagonal boundary matrices. It was suggested in \cite{Cao2013}, see also references thereof, that Baxter's TQ-equation is modified by the insertion of an inhomogeneous term. The construction of the Q-operator may shed more light on that relation and might yield a direct proof of the TQ-equation as presented here for diagonal boundary operators. 

The generalisation of the Q-operator construction to higher rank algebras is outstanding. In particular supersymmetric higher rank algebras with non-compact representations play an important role in the context of the {\small AdS/CFT}-correspondence, see \cite{Zoubos2010} and references therein and thereof.
For generalisations of the closed Heisenberg spin chain we refer the reader to \cite{Bazhanov2010,Frassek2011,Frassek2010}. 

We plan to return to these questions elsewhere.

 \section*{Acknowledgements} 
We like to thank Patrick Dorey, Evgeny Sklyanin and Matthias Staudacher for inspiring discussions as well as Patrick Dorey for comments on the manuscript.
The research leading to these results has received funding from the People Programme
(Marie Curie Actions) of the European Union’s Seventh Framework Programme FP7/2007-
2013/ under REA Grant Agreement No 317089 (GATIS).
\appendix
\section{Crossing symmetry of the transfer matrix}\label{app:tcros}
The crossing symmetry of the transfer matrix 
can be shown using the crossing relation for the R-matrix
\begin{equation}\label{eq:croslax}
 \OO_1R_{12}(-z-1)\OO_1^{-1}=-R_{12}^{t_1}(z)\,,
\end{equation} 
as well as the boundary crossing relations 
\begin{equation}\label{eq:crosl}
\OO_1\KL_1(-z-1)\OO_1^{-1}=\frac{1}{2(z+1)}\tr_2 \perm_{12}R_{12}(2z)\KL_2(z)\,,
\end{equation} 
and
\begin{equation}\label{eq:crosr}
\OO_1\KR_1(-z-1)\OO_1^{-1}=  -\frac{1}{2z}\tr_2 \perm_{12}R_{12}(-2z-2)\KR_2(z)\,.
\end{equation} 
Here the matrix $\OO$ is given by 
\begin{equation}
 \OO=\left(\begin{array}{cc}
              0&1\\
	     -1&0
             \end{array}\right)\,,
\end{equation} 
with $\OO^{t}=\OO^{-1}$.
The subscripts in the equations above denote the corresponding space.
Using the crossing relation \eqref{eq:croslax} and the Yang-Baxter equation \eqref{eq:rrr} we find that \eqref{eq:crosr} implies
\begin{equation}\label{eq:crosu}
\OO_1\mathcal{U}_1(-z-1)\OO_1^{-1}=  -\frac{1}{2z}\tr_2 \perm_{12}R_{12}(-2z-2)\mathcal{M}_2^{t_q}(z)\KR_2(z)\hat{ \mathcal{M}}_2^{t_q}(z)\,,
\end{equation} 
where $\mathcal{U}$ is the double-row monodromy in \eqref{eq:trans} and the $t_q$ the transposition in the quantum space whose subscripts are suppressed. Substituting \eqref{eq:crosl} and \eqref{eq:crosu} into the definition of $\top(-z-1)$ in \eqref{eq:trans} we find
\begin{equation}
 \top(-z-1)=\top(z)\,.
\end{equation} 
Here we used the crossing unitarity relation in \eqref{eq:unitR} and the relation $\mathcal{M}^{t_q}_0(z)=\hat{\mathcal{M}}^{t_0}_0(z) $ with $t_0$ the transposition in the auxiliary space. 
\section{Trace formulae}\label{app:trace}
In this appendix we evaluate the infinite and finite sums for the combination of Gamma functions appearing in Section~\ref{sec:traceoveraux} and \ref{sec:largez}.

In the following we repeatedly use the reflection formula of the Gamma function
\begin{equation}\label{eq:gammareflection}
\Gamma(x) \Gamma(1-x)=\frac{\pi}{\sin(\pi x)} \, ,
\end{equation}
and the generalised hypergeometric function defined as
\begin{equation}\label{eq:hypgeomfunc}
_pF_q (a_1,a_2,\ldots,a_p;b_1,b_2,\ldots,b_q;z)=\sum_{n=0}^\infty\frac{(a_1)_n(a_2)_n\cdots(a_p)_n}{(b_1)_n(b_2)_n\cdots(b_q)_n}\frac{z^n}{n!} \, ,
\end{equation}
where
\begin{equation}\label{eq:pochhammer}
(c)_m=c(c+1)\dots(c+m-1)=\frac{\Gamma(c+m)}{\Gamma(c)}\, ,
\end{equation}
denotes the Pochhammer symbol.

\addtocontents{toc}{\protect\setcounter{tocdepth}{1}}
\subsection{Proof of the trace formula \eqref{eq:sumf}}

In this section we evaluate the trace cf.~\eqref{eq:sumf} which sums over the states of the infinite-dimensional oscillator space
\begin{equation}
\tr \,\frac{\Gamma(x_\pm-\NN)}{\Gamma(y_\pm-\NN)}\,\NN^k
=\sum_{n=0}^\infty \frac{\Gamma(x_\pm-n)}{\Gamma(y_\pm-n)}\,n^k 
\, .
\end{equation}

If $k=0$ using \eqref{eq:gammareflection}, \eqref{eq:hypgeomfunc} and \eqref{eq:pochhammer} we can write the infinite sum as
\begin{equation}
\sum_{n=0}^\infty \frac{\Gamma(x_\pm-n)}{\Gamma(y_\pm-n)}=\frac{\Gamma(x_\pm)}{\Gamma(y_\pm)}{}_{2}F_{1}(1,1-y_\pm;1-x_\pm;1) \, .
\end{equation}
The hypergeometric function $_2F_1$ takes a special value if its argument $z$ equals to $1$ \cite{AbramowitzStegun:1964}
\begin{equation}\label{eq:hypgemz1}
_2F_1(a,b;c;1)=\frac{\Gamma(c)\Gamma(c-b-a)}{\Gamma(c-a)\Gamma(c-b)} \, .
\end{equation} 
This brings us to the result for the $k=0$ case
\begin{equation}
\tr \,\frac{\Gamma(x_\pm-\NN)}{\Gamma(y_\pm-\NN)}=\frac{\Gamma(1+x_\pm)}{(1+x_\pm-y_\pm)\Gamma(y_\pm)}\, .
\end{equation}

If $k\in\mathbb{N}^+$ the first term in the sum is zero, hence by shifting the summation index and using \eqref{eq:gammareflection}, \eqref{eq:hypgeomfunc} and \eqref{eq:pochhammer} the sum takes the form
\begin{equation}\label{eq:qqqq}
\sum_{n=0}^\infty \frac{\Gamma(x_\pm-n)}{\Gamma(y_\pm-n)}n^k=\frac{\Gamma(x_\pm-1)}{\Gamma(y_\pm-1)}{}_{k+1}F_{k}(\underbrace{2,\ldots,2}_k,2-y_\pm;\underbrace{1,\ldots,1}_{k-1},2-x_\pm;1) \, .
\end{equation}
Euler’s integral transform relates a generalised hypergeometric functions of a given order to one of  a lower order via
\begin{equation}
\begin{split}
_{p+1}F_{q+1}(a_1,\ldots,a_{p+1};b_1,\ldots,b_{q+1};z)&=\frac{\Gamma(b_{q+1})}{\Gamma(a_{p+1})\Gamma(b_{q+1}-a_{p+1})}\int_0^1  t^{a_{p+1}-1} (1-t)^{b_{q+1}-a_{p+1}-1}    \\
&\qquad\times{}_{p}F_{q}(a_1,a_2,\ldots,a_{p};b_1,b_2,\ldots,b_{q};tz) \mathrm{d}t \, ,
\end{split}
\end{equation}
In our case we have
\begin{equation}
\begin{split}
{}_{k+1}F_{k}(2,\ldots,2,2-y_\pm;1,\ldots,1,2-x_\pm;1)&=\frac{\Gamma(2-x_\pm)}{\Gamma(2-y_\pm)\Gamma(y_\pm-x_\pm)}\int_0^1 t^{1-y_\pm} (1-t)^{y_\pm-x_\pm-1}    \\
&\qquad\times{}_{k}F_{k-1}(2,\ldots,2;1,\ldots,1;t) \mathrm{d}t \, ,
\end{split}
\end{equation}
cf.~\eqref{eq:qqqq}.
The hypergeometric function with the arguments above is related to the polylogarithm 
\begin{equation}
{}_{k}F_{k-1}(2,\ldots,2;1,\ldots,1;t)=t^{-1} \Li_{-k}(t) \, .
\end{equation}
For negative integers $-k$ the polylogarithm can be written as
\begin{equation}
\Li_{-k}(t)=\sum_{j=0}^{k}j!S_2(k+1,j+1)\left(\frac{t}{1-t}\right)^{j+1} \, ,
\end{equation}
where $S_2$ denotes the Stirling numbers of second kind \cite{AbramowitzStegun:1964}. It is the consequence of the repeated action of the derivative property $t\, \partial_t \Li_n (t)=\Li_{n-1}(t)$ on $\Li_0(t)=\frac{t}{1-t}$.
From the definition of the Beta-function
\begin{equation}
B(x,y)=\int_{0}^{1}t^{x-1}(1-t)^{y-1}dt=\frac{\Gamma(x)\Gamma(y)}{\Gamma(x+y)} \, ,\end{equation}
and the reflection formula \eqref{eq:gammareflection} we obtain the final result 
\begin{equation}
\tr \,\frac{\Gamma(x_\pm-\NN)}{\Gamma(y_\pm-\NN)}\,\NN^k =-\frac{\Gamma(x_\pm)}{\Gamma(y_\pm-x_\pm)}\sum_{j=0}^k(-1)^j\, j!\,S_2(k+1,j+1)\,\frac{\Gamma(y_\pm-x_\pm-1-j)}{\Gamma(y_\pm-1-j)} \,, 
\end{equation}
if $k\in\mathbb{N}^+$.

The result for the $k=0$ and $k\in\mathbb{N}^+$ cases are exactly \eqref{eq:sumf}. 

\subsection{Sums appearing in Section \ref{sec:largez}}

In Section~\ref{sec:largez} two types of sums appear in the derivation of the leading order $z$ behaviour.

The first reads
\begin{eqnarray}\label{eq:GammaSumTrick1}
\sum_{k=0}^{j}\left(-1\right)^{k}\binom{j}{k}\frac{\Gamma\left(A+k\right)}{\Gamma\left(B+k\right)} =
\frac{\Gamma\left(A\right)\Gamma\left(B-A+j\right)}{\Gamma\left(B-A\right)\Gamma\left(B+j\right)}\, ,
\end{eqnarray}
where $A$, $B$ are parameters and $n\in\mathbb{N}$. For the evaluation of the sum we used \eqref{eq:gammareflection}, \eqref{eq:hypgeomfunc}, \eqref{eq:pochhammer}, \eqref{eq:hypgemz1} and also the property of the Pochhammer symbol with negative integer values, namely that $(-a)_k=0$ if $a\in\mathbb{N}$ and $a<k$, cf.~\eqref{eq:pochhammer}.

The second types of sum is 
\begin{eqnarray}\label{eq:GammaSumTrick2}
\sum_{i=0}^{n}\frac{\Gamma\left(A+i\right)}{\Gamma\left(1+i\right)} = \frac{1}{A}\frac{\Gamma\left(1+n+A\right)}{\Gamma\left(1+n\right)}\, ,
\end{eqnarray}
where $A$ is arbitrary parameter and the relation can be shown by induction in $n$.

\bibliographystyle{hieeetr}
\bibliography{qref}{}

\end{document}